# Identifying Hubs in Undergraduate Course Networks Based on Scaled Co-Enrollments: Extended Version

Gary M. Weiss, Nam Nguyen, Karla Dominguez and Daniel D. Leeds
Department of Computer and Information Science
Fordham University, New York, NY
{gaweiss, nnguyen56, kdominguezmelo, dleeds}@fordham.edu

## ABSTRACT
Understanding course enrollment patterns is valuable to predict upcoming demands for future courses, and to provide student with realistic courses to pursue given their current backgrounds. This study uses undergraduate student enrollment data to form networks of courses where connections are based on student co-enrollments. The course networks generated in this paper are based on eight years of undergraduate course enrollment data from a large metropolitan university. The networks are analyzed to identify "hub" courses often taken with many other courses. Two notions of hubs are considered: one focused on raw popularity across all students, and one focused on proportional likelihoods of co-enrollment with other courses. A variety of network metrics are calculated to evaluate the course networks. Academic departments and high-level academic categories, such as Humanities vs STEM, are studied for their influence over course groupings. The identification of hub courses has practical applications, since it can help better predict the impact of changes in course offerings and in course popularity, and in the case of interdisciplinary hub courses, can be used to increase or decrease interest and enrollments in specific academic departments and areas.

## 1. INTRODUCTION

Universities typically offer thousands of different courses across dozens of departments. The interrelationships between courses that are taken together, especially those in different departments, is often not well understood or even studied. This paper addresses this deficiency by forming course networks and then applying network analysis algorithms and metrics to these networks. The course networks are formed by representing courses as nodes and adding edges between each pair of courses that are often taken by the same students. A particular emphasis of this study is on identifying and studying "hub" courses. Hubs are generally defined as nodes in a network that are connected to many other nodes and hence have a high degree count [2]. This study utilizes three popular centrality metrics to identify course hubs and compares the results when using each metric. This study is an extended version of a study published in the International Conference on Educational Data Mining and includes substantially more experiment results [14].

The type of network analysis utilized in this paper has been applied to other domains and has proven to be very useful. Social networks like Facebook have been studied extensively using network analysis [4], and in such cases hubs may correspond to influencers that have an outsized impact on getting others to purchase a product or service. Network analysis has also been applied to the World Wide Web [11] using many of the same network metrics utilized in this paper. Kleinberg's HITS algorithm [7] identifies hubs and authorities on the Web and uses this information to improve web searches. Hubs also play a role in the PageRank algorithm [10], which is the foundation of Google's search algorithm.

Identifying and analyzing hub courses can provide concrete benefits. First, because such courses are heavily associated with other courses, they can be used for better planning of resources, especially if a change is to be made in the frequency or capacity of such courses. Secondly, because of these connections, change can be purposefully made to hub courses to drive (or diminish) student interest in an area or academic discipline. For example, there is currently a need for more STEM (Science, Technology, Engineering, and Math) professionals and a belief that universities should graduate more students in STEM disciplines. If a hub course is well connected to STEM courses, promoting the hub course can lead to increased enrollments in STEM courses—even if the hub course is not a STEM course. Undoubtedly, just like with other descriptive data mining efforts, additional applications for this type of analysis will be discovered over time.

The course network that is analyzed in this study is based on eight years of undergraduate student course enrollment data from a large metropolitan university. An edge connects two courses in the network if the number of students that take both courses is above a specific threshold. Two types of thresholding mechanisms are considered. A static threshold is a fixed value that is the same for all pairs of courses. Because this static threshold is biased toward very popular courses, the study additionally utilizes a dynamic threshold that is set so that only courses that are taken together *relatively* frequently (i.e., relative to their popularity) will be joined by an edge. We shall see that this threshold shifts hub classes from the humanities to STEM disciplines. A variety of popular network statistics are also calculated and analyzed, finding tighter course groupings within STEM and looser groupings within the humanities and social sciences. The analysis in this paper focuses on individual courses, courses grouped by academic department, and courses within six high-level course categories.

## 2. DATASET DESCRIPTION

This section describes the data utilized in this study. Section 2.1 describes the student course enrollment dataset, where each record corresponds to a single student in a single class section. This information is then aggregated and transformed into a course-pair dataset, where there is one record for each pair of courses. The course-pair dataset is used to form the course networks used throughout this paper. This same data set is also used in two of our research group's other studies, one of which analyzes the impact of course sequencing on student grades [6], and the other that forms course networks and analyzes course relationships based on the correlation of grades between courses [8]. This later study performs a somewhat similar analysis to the one provided in this paper, but with a very different notion of course similarity.

## 2.1 Student Course Enrollment Dataset

This dataset contains records that describe individual undergraduate course enrollments at the student level. The dataset is from Fordham university and contains enrollment data from an eight-year period. Table 1 provides an example of such a record and includes identifying information about the student and the course. Note that each course can have multiple sections, which may occur during the same semester or during different semesters. The student's final grade is not used in this study but is used in other ongoing studies concerning instructor effectiveness and the impact of course sequencing. The dataset contains 446,508 records, which cover 24,691 distinct students and 22,608 course sections spanning 83 academic majors.

**Table 1. Sample student course enrollment record**

| Feature name | Example |
|---|---|
| Student ID | S01 |
| Course Title | Computer Science I |
| Department Name | Computer Science |
| Course Number | 1600 |
| Section Number | 1 |
| Semester Code | Fall2016 |
| Final Grade | 4.0 |

## 2.2 Course-Pair Dataset

The course-pair dataset is generated from the student course enrollment dataset by aggregating the information from every pair of courses. Table 2 provides a sample record, which includes identifying information about each of the two courses, the number of students that have taken each course, and the number of common students that have taken both courses. The department associated with each course is also mapped to one of the six major course categories. Note that the course-pair dataset is at the course level, so that it includes data from all sections of the course over the eight-year period. The course-pair dataset contains 78,173 records, which are based on a total of 1,763 distinct courses. The dataset does not contain all possible course pairings, because, as described in Section 4, course pairs with fewer than 20 common students are excluded. The course-pair dataset is generated from the student course enrollment dataset using a publicly available Python-based software tool developed by our research group [12]. The tool also generates the network metrics displayed in the results section.

**Table 2. Sample course-pair record**

| Feature name | Example |
|---|---|
| Course1 | Banned Books |
| Department | English |
| Total students | 10670 |
| Course2 | Film Adaptation |
| Department | Film & TV |
| Total students | 63 |
| Common students | 40 |

## 3. NETWORK ANALYSIS METRICS

The course-pair dataset is used to form course networks by viewing each course as a node and connecting nodes that have a sufficient number of common students, as specified in Section 4. Table 3 summarizes the network analysis metrics that are used to describe the course networks.

The first three metrics in Table 3, density, diameter, and average clustering coefficient [1], are computed based on an entire network or subnetwork. The *density* of a network graph provides an indication of how tightly connected the network is and is computed as the fraction of all possible edges that are present. The *diameter* of a network is the shortest distance between the two most distant nodes in the network that are connected. The *average clustering coefficient (ACC)* is an indicator of the nodes' tendency to form tightly knitted group and is computed by determining the fraction of pairs of nodes connected to a given node that are themselves connected. If we compare a subnetwork associated with the courses offered by a single department to the global course network, we expect the department subnetwork to have a higher density, smaller diameter, and higher average clustering coefficient, because courses within a discipline should be more tightly connected. The results presented later in Table 5 conform to these expectations.

**Table 3. Summary of network analysis metrics**

| Metric | Summary Description | Range |
|---|---|---|
| Density | Fraction of all possible edges that are present. | 0 - 1 |
| Diameter | Maximum distance between any pair of nodes in the network. | $Z^+$ |
| Ave. Clustering Coefficient (ACC) | Fraction of pairs of nodes, which are neighbors of a given node, that are connected to each other. | 0 - 1 |
| Degree Centrality | The number of edges incident on the node (i.e., its degree count). | $Z^+$ |
| Eigenvector centrality | Centrality measure based on the centrality of a node's neighbors. | $\geq 0$ |
| Betweenness centrality | Measure of all the shortest paths that pass through a node | $\geq 0$ |

The last three metrics listed in Table 3 are defined for each *node* in the network and can be used to help identify hubs. They include three centrality measures: degree centrality, eigenvector centrality, and betweenness centrality [9]. The *degree centrality* of a node is the number of edges incident on the node, which is the degree of the node [3]. *Eigenvector centrality* is an extension of the degree centrality that considers not only the connectedness of the node, but the importance of its neighbors [13]. *Betweenness centrality* captures how often a given node is located on the shortest path between two random points in the network [5], and rewards nodes that serve as a bridge between disparate parts of the network. High degree count is not sufficient to guarantee high betweenness centrality. Each of these three centrality measures can be used to identify a different type of hub course. For our analyses, a rank is computed for each of these centrality metrics, where a rank of 1 identifies the node with the highest centrality. These ranks are used in Tables 6 and 7 to order our results. We also compute a *combined rank* as the median value of the three centrality ranks.

## 4. EDGE INCLUSION METHODOLOGY

Forming the course networks is a relatively simple process. Each course is represented by a node and an edge is added between two nodes if the courses, aggregated over all course sections, have a sufficient number of common students. The key issue is how to determine the "sufficient number" of common students. This section focuses on that decision by defining static and dynamic thresholding mechanisms and providing guidance on how the specific threshold values were established.

### 4.1 Static Thresholding of Edges

The static threshold is based on the number of common students between two courses, independent of how many students take each course. The course network will differ greatly based on the threshold value that is selected. To gain insight into the impact of specific threshold values, Figure 1 shows the distribution of common students by course pair. Each bin in the figure includes course-pairs with a range of common students. The distribution shows that most course-pairs have very few common students. The orange curve is a cumulative curve that corresponds to the y-axis values listed to the right (varying between 0% and 80%) and represents the percentage of course-pairs that are *maintained* for each possible common student threshold value. For example, a threshold value of 20 maintains 11% of all course-pairs with at least one student in common (i.e., 89% are discarded). This value is the one that is utilized in this study for our static threshold. While excluding 89% of the course pairs may seem extreme, most courses are rarely taken by the same set of students. This should not be too surprising since very few students will take upper-level courses in disparate disciplines. Other static threshold values will be evaluated in future studies, but the dynamic threshold described next already provides a more restrictive alternative.

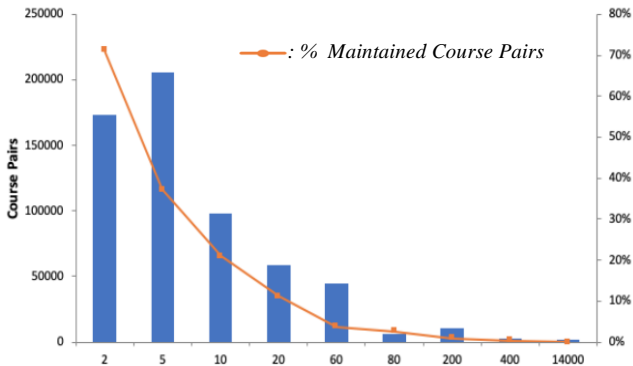

**Figure 1. Number of students in common**

### 4.2 Dynamic Thresholding of Edges

The static thresholding mechanism is heavily biased towards popular courses that are taken very frequently. Such courses are likely to have an edge to many other courses, even if only a small fraction of the students in the popular course take specific other courses. Courses associated with a Fordham's core curriculum requirement will be very popular, have many edges, and therefore be identified as hub courses. This may be the desired behavior in some cases, but we also consider a dynamic threshold that relies primarily on the *co-occurrence rate* of courses, and hence identifies another type of hub.

The co-occurrence threshold rate is represented by a constant $k$ and the dynamic threshold is determined by multiplying this rate by the number of students in the course within the course-pair with the most students. However, because we still want to make sure there is at least some minimum number of students in common, a static threshold is used as the floor for the dynamic threshold. For this study we use the static threshold of 20 as the floor of the dynamic threshold. The dynamic threshold, *d-thresh*, associated with two courses, $C_1$ and $C_2$, is provided in Equation 1, where $C_x$.students represents the number of students who have taken class $C_x$.

$$d\text{-}thresh(C_1, C_2) = \max(20, k \times \max(C_1.\text{students}, C_2.\text{students})) \quad [1]$$

The dynamic threshold is heavily dependent on the co-occurrence rate $k$. To help set this value appropriately, Figure 2 shows the distribution of course-pairs for each co-occurrence rate, for the course pairs that satisfy the static threshold of 20. The co-occurrence rate is the number of common students divided by the number of students in the course with more students. The co-occurrence rate distribution in Figure 2 is heavily skewed to the smaller values, just as the number of common students was skewed to the smaller values in Figure 1. The bar at the far right at x=1.0 is associated with course pairs with the same course in both positions and should be ignored. After some experimentation we decided on a co-occurrence rate threshold $k = 0.017$, which is the value that leads to the most stable centrality measures while excluding the fewest number of edges. Based on the cumulative curve shown in Figure 2, which represents the fraction of edges discarded, this value of $k$ discards 39% of the edges that satisfy the static threshold.

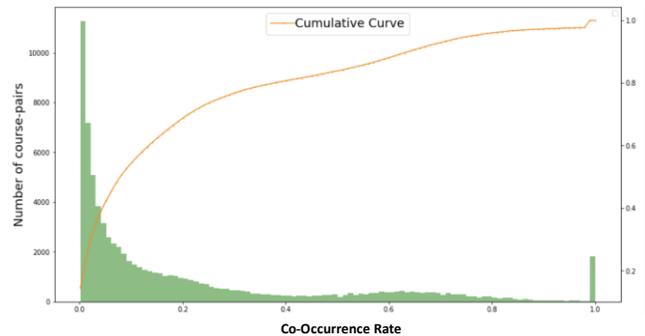

**Figure 2. Co-Occurrence Rate Distribution**

To illustrate how the dynamic threshold works, we apply it to the course "Art History Seminar", which has a total of 123 students. As shown in Table 4 (column 2), there are 22 courses that share at least 20 students in common with this course and therefore satisfy the static threshold. However, 9 courses have fewer common students than the computed dynamic threshold and hence are pruned by the dynamic threshold (pruned edges are denoted using boldface). As anticipated, the courses affected by the dynamic threshold are those with a very large number of students, as indicated by the values in the third column. The smallest course that is impacted is "Intro to World Art History," which has 2,267 students. In this example, every course that satisfies the static threshold but is pruned by the dynamic threshold satisfies one of the universities' core curriculum requirements. Without the dynamic threshold all core courses would likely be kept simply due to their popularity.

**Table 4. Dynamic threshold for Art History seminar course**

| Course2 | Common Students | Students Course2 | Dynamic Threshold |
|---|---|---|---|
| **Intro Cultural Anthro.** | 23 | 2514 | 43 |
| **Intro World Art History** | 36 | 2267 | 39 |
| Ancient American Art | 21 | 34 | 20 |
| Renaissance Portraits | 25 | 65 | 20 |
| 17th Century Art | 22 | 47 | 20 |
| 19th Century Art | 28 | 60 | 20 |
| 20th Century Art | 43 | 130 | 20 |
| Museum Methods | 21 | 49 | 20 |
| Age of Cathedrals | 20 | 39 | 20 |
| Michelangelo | 23 | 57 | 20 |
| Aztec Art | 22 | 61 | 20 |
| Senior Seminar | 123 | 123 | 20 |
| **Composition II** | 58 | 12446 | 211 |
| **Banned Books** | 47 | 10670 | 181 |
| Intermed. French II | 20 | 1329 | 23 |
| French Lang. and Lit. | 26 | 1460 | 25 |
| **Finite Math** | 42 | 4976 | 85 |
| **Philos. of Human Nature** | 56 | 12972 | 220 |
| **Philosophical Ethics** | 58 | 11218 | 191 |
| **Spanish Lang. & Lit.** | 27 | 3738 | 64 |
| **Faith & Critical Reason** | 56 | 13317 | 226 |
| Visual Thinking | 64 | 821 | 20 |

## 5. RESULTS

This section describes our results for the course networks in terms of the network analysis metrics presented earlier and more in-depth analyses related to hub courses. Results are presented for both the static and dynamic thresholds and each of the three centrality metrics. The hub results are presented and analyzed for individual courses, courses grouped by academic department, and courses grouped by course category. This study utilizes the following six course categories: Arts, Communication and Media Studies, Humanities, Modern Languages, Social Sciences, and STEM. The mapping from academic department to course category is provided in Table 5.

### 5.1 Network Metric Results

Table 5 presents the values of the previously defined network metrics for the course network and subnetworks at the department and category levels. The course categories are denoted in bold and the departments associated with that category are listed right after the category. The value as the category level reflects the median values across the member departments. The first row of data provides the values over all courses in the course network. The color of the cells reflects the magnitude of the cell value, with red (green) used for the highest (lowest) values. The colors for the departments and categories are determined independently.

The network covering all courses has a high diameter and low density compared to the subnetworks, which is expected since it includes many diverse courses that will generally be only loosely connected. Courses associated with a specific department are most likely associated with a major and students within the major will take many of these courses. The dynamic threshold, which will result in fewer edges than the static threshold, decreases the density, average clustering coefficient (ACC), and number of edges, while increasing the diameter.

Study of departmental subnetworks shows dynamic thresholding most dramatically decreases edges for Philosophy (52% decrease from 214 to 102), English (44% decrease from 462 to 258), and Theology (35% decrease from 179 to 116), which are major academic fields of study that include many core curriculum courses. This drop is mirrored by ACC. Conversely, the diameter maintains similar values for most departments, regardless of the threshold applied. Overall, dynamic thresholding has a substantial impact on density and ACC of Humanities and Social Science courses, and only a minimal impact on other categories, likely reflecting our university core curriculum's emphasis on humanities and social science courses.

It is interesting to note that for both static and dynamic thresholds, the STEM courses have much higher density and form much more dense clusters (based on ACC) than humanities courses. This indicates that humanities students are less likely to take the same group of courses in their discipline. This behavior is consistent with our understanding that humanities majors have fewer required courses than STEM majors. Humanities departments have the highest number of nodes (distinct courses taken), closely followed by Social Science, suggesting that those disciplines allow more flexibility in the choices provided to the students. The Modern Languages category also has a relatively high density and ACC; language courses, like science courses, typically rely on prerequisite course requirements to ensure students have proper preparation.

### 5.2 Hub Analysis

Hubs play a special role in network structures and play an important role in understanding and utilizing the information in course co-enrollment networks. In Section 3 we introduced three centrality metrics that are often used to identify hubs: degree centrality, eigenvector centrality and betweenness centrality. These metrics are applied to every course in our course network and the results are described and analyzed in this section.

Table 6 lists the top-20 hub courses for each of the three centrality metrics, when using both the static and dynamic thresholds to form the network. The first few entries for the static threshold vary only slightly depending on which of the three centrality metrics is used, and in each case the course that is listed corresponds to a core curriculum requirement that can be satisfied only by a single course. Most of the remaining courses in the top 20 also satisfy a core requirement, but they are not as popular because the requirement can be satisfied by any of several courses. Very few STEM courses are listed and the ones that are listed are introductory and satisfy a core requirement (e.g., "Finite Mathematics"). Thus, we see that the hubs identified using the static threshold are based on raw popularity. Most of the courses identified using the dynamic threshold also satisfy a core requirement, but often there are many courses that can satisfy the core requirement. In addition, the centrality metrics tend to identify different hub courses—there is much more diversity between the metrics when using the dynamic threshold versus the static threshold. This makes sense since the metrics do measure different aspects of centrality, whereas in the static cases those differences are overwhelmed by the overall popularity of the courses.

Table 5. Summary course network statistics based on category and department

| Category/ Department | Nodes | Static Threshold | | | | Dynamic Threshold | | | |
|---|---|---|---|---|---|---|---|---|---|
| | | Edges | Density | Diam. | ACC | Edges | Density | Diam. | ACC |
| **ALL** | 1763 | 39968 | 0.03 | 4 | 0.74 | 24323 | 0.02 | 6 | 0.40 |
| **Arts** | **41.5** | **239** | **0.32** | **3** | **0.56** | **231** | **0.29** | **2.5** | **0.56** |
| Dance | 54 | 1236 | 0.86 | 3 | 0.95 | 1236 | 0.86 | 3 | 0.95 |
| Music | 24 | 87 | 0.32 | 3 | 0.51 | 73 | 0.26 | 2 | 0.52 |
| Theatre | 47 | 359 | 0.33 | 3 | 0.61 | 347 | 0.32 | 4 | 0.59 |
| Visual Arts | 36 | 119 | 0.19 | 3 | 0.51 | 114 | 0.18 | 2 | 0.51 |
| **Comm and Media Studies** | **24** | **25** | **0.20** | **2** | **0.16** | **25** | **0.19** | **2** | **0.16** |
| Comm and Media Studies | 94 | 862 | 0.20 | 3 | 0.72 | 828 | 0.19 | 4 | 0.58 |
| Comm & Culture | 32 | 50 | 0.10 | 2 | 0.16 | 50 | 0.10 | 2 | 0.16 |
| Digital Tech & Emerging Media | 15 | 22 | 0.21 | 2 | 0.23 | 22 | 0.21 | 2 | 0.23 |
| Film & TV | 24 | 25 | 0.09 | 1 | 0.00 | 25 | 0.09 | 1 | 0.00 |
| New Media & Digital Design | 6 | 8 | 0.53 | 2 | 0.00 | 8 | 0.53 | 2 | 0.00 |
| **Humanities** | **81** | **179** | **0.06** | **3** | **0.21** | **104** | **0.04** | **3** | **0.08** |
| African & African Amer Studies | 28 | 34 | 0.09 | 2 | 0.11 | 34 | 0.09 | 2 | 0.11 |
| Anthropology | 50 | 122 | 0.10 | 3 | 0.39 | 104 | 0.08 | 3 | 0.29 |
| Art History | 37 | 58 | 0.09 | 3 | 0.21 | 49 | 0.07 | 5 | 0.00 |
| English | 167 | 462 | 0.03 | 3 | 0.59 | 258 | 0.02 | 3 | 0.12 |
| History | 180 | 292 | 0.02 | 3 | 0.17 | 202 | 0.01 | 4 | 0.03 |
| Philosophy | 97 | 214 | 0.05 | 3 | 0.50 | 102 | 0.02 | 2 | 0.02 |
| Theology | 81 | 179 | 0.06 | 2 | 0.21 | 116 | 0.04 | 4 | 0.08 |
| **Modern Languages** | **9** | **19** | **0.53** | **2** | **0.49** | **19** | **0.53** | **2** | **0.38** |
| French | 22 | 60 | 0.26 | 3 | 0.49 | 55 | 0.24 | 2 | 0.37 |
| German | 7 | 15 | 0.71 | 2 | 0.62 | 15 | 0.71 | 2 | 0.62 |
| Greek | 4 | 6 | 1.00 | 2 | 0.00 | 6 | 1.00 | 2 | 0.00 |
| Italian | 13 | 28 | 0.36 | 2 | 0.38 | 28 | 0.36 | 2 | 0.38 |
| Latin | 7 | 13 | 0.62 | 1 | 0.57 | 13 | 0.62 | 1 | 0.57 |
| Mandarin Chinese | 9 | 19 | 0.53 | 1 | 0.56 | 19 | 0.53 | 1 | 0.56 |
| Spanish | 40 | 118 | 0.15 | 3 | 0.49 | 98 | 0.13 | 2 | 0.33 |
| **STEM** | **34** | **295** | **0.47** | **3** | **0.76** | **288** | **0.45** | **3** | **0.75** |
| Biological Sciences | 30 | 274 | 0.63 | 2 | 0.77 | 274 | 0.63 | 2 | 0.77 |
| Chemistry | 28 | 303 | 0.80 | 3 | 0.84 | 299 | 0.79 | 2 | 0.84 |
| Computer Science | 44 | 438 | 0.46 | 3 | 0.75 | 423 | 0.45 | 3 | 0.73 |
| Mathematics | 30 | 210 | 0.48 | 3 | 0.75 | 187 | 0.43 | 3 | 0.74 |
| Natural Science | 54 | 639 | 0.45 | 3 | 0.83 | 639 | 0.45 | 3 | 0.83 |
| Physics | 38 | 286 | 0.41 | 4 | 0.73 | 277 | 0.39 | 3 | 0.74 |
| **Social Science** | **74** | **329** | **0.18** | **2.5** | **0.51** | **285** | **0.16** | **2.5** | **0.44** |
| Economics | 45 | 325 | 0.33 | 2 | 0.64 | 270 | 0.27 | 2 | 0.59 |
| Political Science | 120 | 332 | 0.05 | 3 | 0.35 | 301 | 0.04 | 3 | 0.32 |
| Psychology | 58 | 500 | 0.30 | 2 | 0.69 | 486 | 0.29 | 2 | 0.68 |
| Sociology | 90 | 236 | 0.06 | 3 | 0.37 | 206 | 0.05 | 3 | 0.30 |

Table 6. Top 20 static and dynamic hub courses by each centrality metric

| | Static threshold | | | Dynamic threshold | | |
|---|---|---|---|---|---|---|
| Rank | Degree centrality | Betweenness centrality | Eigenvector centrality | Degree centrality | Betweenness centrality | Eigenvector centrality |
| 1 | Philosophical Ethics | Philosophical Ethics | Faith & Critical Reason | Intro to Phys Anthro | English Lit. Theory | Intro Cultural Anthro. |
| 2 | Faith & Critical Reason | Faith & Critical Reason | Philosophical Ethics | Latin American Hist | Phys Sci: Today's World | Intro World Art History |
| 3 | Philos. of Human Nature | Philos. of Human Nature | Philos. of Human Nature | Foundations of Psych. | Intro to Spanish I | Foundations of Psych. |
| 4 | Composition II | Banned Books | Composition II | Phys Sci: Today's World | Multivariable Calc I | Intro to Sociology |
| 5 | Banned Books | Composition II | Banned Books | Intro World Art History | Intro to Internat. Politics | Latin Amer. History |
| 6 | Finite Mathematics | Spanish Lang & Lit | Basic Macroeconomics | Intro Cultural Anthro. | Latin Amer. History | Intro to Phys Anthro |
| 7 | Spanish Lang & Lit | Finite Mathematics | Finite Mathematics | Calculus II | Films of Moral Struggle | West. Music Traditions |
| 8 | Basic Macroeconomics | Basic Macroeconomics | Spanish Lang & Lit | Films of Moral Struggle | Approaches to Lit | Philosophical Ethics |
| 9 | Intermediate Spanish II | Intermediate Spanish II | Understand. Amer Hist | Physics I Lab | Intro to Phys Anthro | Invitation to Theatre |
| 10 | Basic Microeconomics | Intro to Astronomy | Basic Microeconomics | Intro to Politics | Intro to Politics | Biopsychology |
| 11 | Understand. Amer Hist | Basic Microeconomics | Intro to Sociology | Physics II Lab | Human Function/Dysfun. | Banned Books |
| 12 | Intro to Astronomy | Understanding Amer Hist | Invitation to Theatre | Intro Internat. Politics | Visual Thinking | Structures of Comp Sci |
| 13 | Invitation to Theatre | Intermediate Spanish I | Intermediate Spanish II | Intro Comm. & Med Stud | French Lang & Lit | Calculus I |
| 14 | Intro to Sociology | Invitation to Theatre | Modern Europe | French Lang & Lit | Phys Sci:Past to Present | Urbanism |
| 15 | Intermediate Spanish I | Introduction to Sociology | Statistics I | General Chem Lab II | Intro to French I | Intro to Politics |
| 16 | Modern Europe | Intro to Cultural Anthro | Intermediate Spanish I | Structures of Comp Sci | Honors: History | Spanish Lang & Lit |
| 17 | Statistics I | Statistics I | Western Music Traditions | General Chem Lab I | Intro Comm. & Med Stud | Intro Old Testament |
| 18 | Western Music Traditions | Modern Europe | Intro Cultural Anthro. | Calculus I | Sociological Theory | Applied Calculus I |
| 19 | Intro Cultural Anthro. | Intro to Politics | Intro to Astronomy | General Chem II | Calculus II | Composition I |
| 20 | Intro to Politics | Intro to Media Industries | Intro to Politics | Biopsychology | Foundations of Psych | Personality |

Additional information about the course hubs is provided in Table 7, which identifies the top-20 hubs using the median of the ranks of the three centrality metrics, "Combined Rank". The top half of the table provides the top-20 hubs when using the static threshold, while the bottom half provides the top-20 for the dynamic threshold. Note that the best combined rank when using the dynamic threshold is 3—this occurs because no course consistently ranks above third on all centrality metrics. Unlike Table 6, Table 7 also provides the actual value for each metric. While only the combined rank for the static (dynamic) threshold is used to select the entries in the top (bottom) half of the table, both combined ranks are provided to make it easy to compare the differences between the static and dynamic thresholding mechanisms. These columns clearly show that courses have very different ranks for the static and dynamic thresholds.

The entries in Table 7 overlap heavily with those in Table 6 and we will not repeat the previous observations. However, Table 7 makes it much easier to see the differences between the different centrality metrics. Note that when using the static threshold, the course ranks are quite consistent across all three centrality metrics. This ensures that the combined rank is also highly correlated with each of the individual metrics, and that the degree centrality is usually equal to the combined rank. This correlation is weaker when examining the dynamic threshold, and there are cases when the degree centrality differs substantially from the combined dynamic rank, such as for "Calculus II" and "Physics I Lab" (in these cases the two rank values are underlined and displayed in bold). Nonetheless, degree centrality is still generally close to the combined rank and is identical to it in 12 of the first 20 cases. For this reason, for the rest of our analyses we focus on degree centrality as our metric for identifying hubs. This is attractive since degree centrality is the simplest and most common metric for identifying hubs (recall that degree centrality is based solely on the number of edges connected to a node). More specifically, we utilize a degree count threshold of 200 to identify hub courses. This retains all entries in Table 7 since all courses have degree count of at least 245, and the underlying data ensures that a degree count of 200 will retain the top fifty courses associated with each underlying metric).

Table 7 also clearly shows that the dynamic threshold discounts some well-connected courses, since when using the dynamic threshold no course has a degree count above 287. The table further shows that the other two centrality values, betweenness centrality and eigenvector centrality, are much lower when using the dynamic threshold. Finally, by comparing the combined rank for the static and dynamic thresholds we see that there are very substantial differences. For example, "Philosophical Ethics" is ranked first when using the static threshold but $45^{th}$ when using the dynamic threshold. This large difference indicates that most of the connections to other courses are incidental and simply due to so many students taking the "Philosophical Ethics." There are only two courses that appear in both top-20 lists: "Introduction to Cultural Anthropology" (18 vs. 6) and "Introduction to Politics" (20 vs. 10). These courses are hubs in both the static and dynamic sense: they are relatively popular, and many other courses are taken by the same students at a rate greater than expected if the courses were completely independent.

Table 7. Top-20 static and dynamic course hubs

| Rank | Courses | Combined Rank | | Centrality Rank | | | Centrality Actual value | | |
|---|---|---|---|---|---|---|---|---|---|
| | | Static | Dynamic | Degree | Betweenness | Eigen | Degree | Betweenness | Eigen |
| | **Static Threshold: Top Hubs** | | | | | | | | |
| 1 | Philosophical Ethics | 1 | 45 | 1 | 1 | 2 | 1388 | 0.1465 | 0.1305 |
| 2 | Faith & Critical Reason | 2 | 76 | 2 | 2 | 1 | 1355 | 0.1074 | 0.1306 |
| 3 | Philosophy of Human Nature | 3 | 75 | 3 | 3 | 3 | 1335 | 0.0968 | 0.1304 |
| 4 | Composition II | 4 | 78 | 4 | 5 | 4 | 1310 | 0.0890 | 0.1300 |
| 5 | Banned Books | 5 | 49 | 5 | 4 | 5 | 1307 | 0.0911 | 0.1298 |
| 6 | Finite Mathematics | 7 | 56 | 6 | 7 | 7 | 780 | 0.0158 | 0.1109 |
| 7 | Spanish Lang and Lit | 7 | 29 | 7 | 6 | 8 | 759 | 0.0196 | 0.1102 |
| 8 | Basic Macroeconomics | 8 | 96 | 8 | 8 | 6 | 701 | 0.0109 | 0.1114 |
| 9 | Intermediate Spanish II | 9 | 39 | 9 | 9 | 13 | 666 | 0.0102 | 0.1067 |
| 10 | Basic Microeconomics | 10 | 103 | 10 | 11 | 10 | 658 | 0.0084 | 0.1090 |
| 11 | Understanding American Hist. | 11 | 71 | 11 | 12 | 9 | 652 | 0.0072 | 0.1093 |
| 12 | Intro to Astronomy | 12 | 33 | 12 | 10 | 19 | 629 | 0.0091 | 0.1001 |
| 13 | Invitation to Theatre | 13 | 36 | 13 | 14 | 12 | 629 | 0.0067 | 0.1069 |
| 14 | Intro to Sociology | 14 | 27 | 14 | 15 | 11 | 620 | 0.0065 | 0.1073 |
| 15 | Intermediate Spanish I | 15 | 57 | 15 | 13 | 16 | 608 | 0.0070 | 0.1039 |
| 16 | Modern Europe | 16 | 69 | 16 | 18 | 14 | 605 | 0.0055 | 0.1065 |
| 17 | Statistics I | 17 | 107 | 17 | 17 | 15 | 587 | 0.0055 | 0.1045 |
| 18 | Western Music Traditions | 18 | 31 | 18 | 21 | 17 | 567 | 0.0047 | 0.1030 |
| 19 | Intro to Cultural Anthro | 18 | 6 | 19 | 16 | 18 | 552 | 0.0057 | 0.1012 |
| 20 | Intro to Politics | 20 | 10 | 20 | 19 | 20 | 524 | 0.0051 | 0.0969 |
| | **Dynamic Threshold: Top Hubs** | | | | | | | | |
| 1 | Biopsychology | 31 | 3 | 3 | 20 | 3 | 285 | 0.0063 | 0.0970 |
| 2 | Phys. Science: Today's World | 30 | 4 | 4 | 2 | 63 | 283 | 0.0195 | 0.0814 |
| 3 | Latin American History | 44 | 5 | 2 | 6 | 5 | 287 | 0.0122 | 0.0957 |
| 4 | Intro World Art History | 22 | 5 | 5 | 26 | 2 | 272 | 0.0042 | 0.0984 |
| 5 | Intro to Physical Anthropology | 41 | 6 | 1 | 9 | 6 | 287 | 0.0094 | 0.0957 |
| 6 | Intro to Cultural Anthropology | 18 | 6 | 6 | 33 | 1 | 269 | 0.0037 | 0.0987 |
| 7 | Films of Moral Struggle | 55 | 8 | 8 | 7 | 54 | 259 | 0.0116 | 0.0853 |
| 8 | Intro to Politics | 20 | 10 | 10 | 10 | 15 | 257 | 0.0093 | 0.0911 |
| 9 | Intro to International Politics | 60 | 12 | 12 | 5 | 60 | 255 | 0.0128 | 0.0826 |
| 10 | French Language & Literature | 34 | 14 | 14 | 13 | 23 | 252 | 0.0079 | 0.0891 |
| 11 | Structures of Computer Science | 23 | 16 | 16 | 64 | 12 | 249 | 0.0023 | 0.0928 |
| 13 | Intro. Comm. & Media Studies | 21 | 17 | 13 | 17 | 35 | 254 | 0.0068 | 0.0880 |
| 14 | Calculus I | 29 | 18 | 18 | 49 | 13 | 248 | 0.0029 | 0.0919 |
| 15 | Calculus II | 53 | **19** | **7** | 19 | 32 | 262 | 0.0066 | 0.0883 |
| 16 | Foundations of Psychology | 28 | 20 | 20 | 56 | 10 | 247 | 0.0025 | 0.0942 |
| 17 | Statistics | 50 | 22 | 21 | 30 | 22 | 247 | 0.0038 | 0.0892 |
| 18 | Physics I Lab | 46 | **23** | **9** | 23 | 44 | 259 | 0.0047 | 0.0870 |
| 19 | General Chemistry Lab I | 30 | **24** | **17** | 53 | 24 | 248 | 0.0026 | 0.0891 |
| 20 | Intro to Media Industries | 24 | 24 | 24 | 21 | 43 | 245 | 0.0051 | 0.0870 |

**Table 8. Percent distribution of hub edge linkage by course category (hubs with degree ≥200) with edge info**

| Category | Static threshold | | | | | | | | Dynamic threshold | | | | | | | |
|---|---|---|---|---|---|---|---|---|---|---|---|---|---|---|---|---|
| | Arts | Comm | Hum | Lang | STEM | SocSci | #Edges | %Edges | Arts | Comm | Hum | Lang | STEM | SocSci | #Edges | %Edges |
| Arts | 5 | 14 | 27 | 7 | 26 | 22 | 1520 | 5 | 5 | 9 | 21 | 10 | 36 | 20 | 650 | 5 |
| Comm. | 11 | 24 | 22 | 7 | 20 | 15 | 1426 | 5 | 8 | 27 | 19 | 9 | 21 | 15 | 954 | 7 |
| Hum | 10 | 12 | 31 | 6 | 21 | 20 | 10892 | 35 | 6 | 11 | 20 | 10 | 33 | 21 | 2758 | 19 |
| Lang. | 10 | 16 | 28 | 5 | 18 | 23 | 3219 | 10 | 9 | 17 | 23 | 5 | 22 | 24 | 1773 | 12 |
| STEM | 5 | 8 | 27 | 7 | 32 | 20 | 8479 | 27 | 5 | 6 | 24 | 8 | 36 | 21 | 5543 | 39 |
| SocSci. | 7 | 10 | 25 | 8 | 25 | 25 | 5543 | 18 | 3 | 6 | 23 | 10 | 29 | 29 | 2717 | 19 |

We are interested in the role that hub courses play in connecting courses across course categories. Table 8 shows the distribution of hub edges between the six course categories using a degree centrality threshold of 200. The table displays the percentage of total hub edges from one category (row) to both hub and non-hub courses in another category (column), for both the static and dynamic thresholds. The percentage of total edges, as well as the actual number of edges, associated with each category (row), are also provided. A color scale is applied to the rows to highlight where the hub connections are directed (red is high percentage and green low percentage). As an example, the left half of the first row in Table 8 indicates that using a static threshold, 5% of all Arts courses are connected to other Arts courses and 14% are connected to Communication courses. In addition, the Arts courses have a total of 1,520 edges that comprise 5% of all edges in the course network.

In Table 8, for both thresholds, the categories with the most hub edges are Humanities, STEM, and Social Sciences, while the Arts, Communications, and Modern Language categories have far fewer and consequently their cells in the "%Edges" column are colored in green. A key difference, however, is that the static threshold associates more edges with humanities courses than STEM courses (35% to 27%), whereas the dynamic threshold reverses this trend, generating more than twice as many STEM edges (39% versus 19%). Most core curriculum requirements are associated with humanities courses, and the dynamic threshold has an outsized impact removing courses that are hubs simply due to their popularity.

Table 8 most emphasizes hub links to humanities courses under the static threshold, consistent with the fact many common core courses are in the Humanities category. However, when using the dynamic threshold this shifts dramatically and now STEM courses are linked more often than any other category—predominantly driven by STEM having the majority of edges under the dynamic threshold. It is particularly interesting that more edges link from the Humanities to STEM courses than to humanities courses. Examining the underlying data, we find that the humanities courses "Introduction to Cultural Anthropology", "Introduction to Physical Anthropology", and "Introduction to Art History" all connect to STEM hub courses. Most connections for courses in the Anthropology and Art History departments go towards the Biological Sciences and Natural Science departments. While "Introduction to Physical Anthropology" is part of the Natural Science major requirement, it also satisfies the Life Science core curriculum requirement for non-Science majors, so it is interesting to see that it was popular with students from Science departments. This course specifically serves as a general survey of the biological focus of Anthropology.

Also notable, Communications and Social Sciences have more links to themselves than to any other category for both static and dynamic thresholds, even though these categories do not have as many total links as other categories. In comparison, Humanities and STEM have many self-directed edges, but this may be due to the large number of total edges in each category. The Languages category have mostly internal links, and an intermediate number of edges overall for both thresholds.

Social Science hubs have a significant number of connections towards STEM courses in Table 8, commensurate to connections back towards Social Science. Most of the connections to STEM refer to courses in Biological Sciences, particularly from the Psychology course "Foundations of Psychology." This course is a main requirement for the Psychology major but is not part of the common core curriculum. This course also has a significant number of connections with the Natural Science department. Overall, the number of connections from non-STEM courses to STEM courses when using the dynamic threshold is a bit of a surprise. Conversely, STEM hubs made many connections to the Social Science category in Table 8; these connections are largely directed towards the Economics department, which requires a strong mathematical base.

## 6. CONCLUSIONS

This study formed course network graphs using eight years of undergraduate course-grade data from Fordham university, and then analyzed the resulting network. The network, metrics, and various summary tables were generated using a publicly available Python-based tool that was created by our research group and is available to other researchers [12]. A special focus of the research was in identifying hub courses that are connected to many other courses based on student co-enrollments. Edges were formed using both a static threshold and a dynamic threshold; the dynamic threshold focused more on the rates of co-enrollments than the absolute number of co-enrollments, thus identifying two very different types of hubs. Our analysis examined the results at the level of individual courses, courses grouped by academic department, and higher-level course categories. This work provides important insights on relations among classes as well as import insights on the metrics that would be most naturally applied to characterize these relations. This study is an extended version of an existing study [14].

All three common network centrality metrics (degree centrality, betweenness centrality, and eigenvector centrality) identify a similar set of hub courses when the static thresholding mechanism is used to determine the edges in the network. However, the three metrics behave much less similarly when dynamic thresholding is used, indicating that one must carefully consider which metric to employ. Nonetheless, degree centrality still yields a reasonable approximation of the other two metrics when using dynamic thresholding, supporting use of degree centrality as the primary centrality metric for studying course co-enrollment networks.

The static and dynamic thresholds yield very different course networks and hubs. The static threshold places more emphasis on raw course popularity, and the very top-ranked hubs correspond to courses that uniquely satisfy a core requirement. The dynamic threshold partially helps avoid this popularity bias; however, popular courses are favored by both thresholds. Due to the many mandatory core courses in Humanities and the variety of core options in STEM, the dynamic threshold substantially shifted apparent hub focus from Humanities to STEM. Future analyses of course relations and discipline relations must continue to carefully weigh the influence of popularity or the mandatory nature of certain courses.

For both static and dynamic thresholding, the STEM courses have the highest density and form tightly connected clusters, while humanities courses have the opposite behavior; this is likely due to the more extensive use of prerequisites in STEM disciplines.

Our analysis also identified relatively large numbers of edges between the different course categories. Edge distributions shifted between thresholds, favoring humanities for the static threshold and STEM for the dynamic threshold. Study of courses forming individual edges provided additional insights. The strong connection between humanities and STEM courses was driven by humanities courses like "Introduction to Physical Anthropology," which has a strong STEM component; the connection between social sciences and STEM was driven by courses like "Foundations of Psychology" which is linked to STEM courses in Biology (Psychology students must take several biology courses).

The work described in this paper provides a better understanding of course co-enrollment patterns, suggesting directions for valuable practical applications. Strong models of co-enrollment patterns can help with course planning and ensuring sufficient numbers of course sections are offered. The connections that occur when using the dynamic thresholds shows a potential causal relationship between course enrollments, since the edges are not due simply due to course popularity, and these relationships can be used to drive interest in courses and disciplines. At the most granular level this was perhaps already known, but the course networks reveal additional detail and more quantitative relationships. For example, as we saw, Psychology majors must take Biology courses, which can potentially lead to even further interest in biology and could lead to some students changing their majors. Perhaps more interesting cases occur when the linkage between subject matter is not necessarily as obvious, as was the relationship between anthropology and STEM courses. We view this work as a foundational step in better understanding course co-enrollments and expect that over time new applications for these networks, and networks statistics, will be discovered.

There are many ways in which this work can be extended and improved. The thresholding mechanisms can be further investigated and improved. Our analysis can be repeated with larger values for the static threshold, which was set at 20 for this study. The dynamic threshold can also be refined to based more on the underlying probabilities of each course being taken, so edges are included only between courses where the co-occurrence is substantially more likely than by chance. We can also look at better clustering of courses and consider clustering courses in areas of the graph with a high clustering coefficient. Additionally, it would be interesting to incorporate course ordering information and extend our metrics and analyses to cover the resulting directed graphs. Though more time would be needed to parse through the results, we would like to perform a more thorough hub analysis at the department level. While aggregating the courses by category was informative, the category-level patterns may be dominated by the more popular departments. Finally, although this is a descriptive data mining task, some validation can still be done. By partitioning the underlying student enrollment records into distinct subsets, perhaps by year, we can build our network using only part of the data and verify that the observed conclusions hold on the remaining "test" data.

# 7. REFERENCES

[1] Arif, T. 2015. Mining and Analyzing Academic Social Networks, *International Journal of Computer Applications Technology and Research*, 4, 878-883. 10.7753/IJCATR0412.1001.

[2] Barabási, A. 2016. *Network Science*, Cambridge University Press.

[3] Bonacich, P. 1987. Power and centrality: A family of measures. *American Journal of Sociology*, 92(5), 1170-1182.

[4] Catanese, S. A., De Meo, P., Ferrara, E., Fiumara, G., and Provetti, A. 2011. Crawling Facebook for social network analysis purposes, *Proceedings of the International Conference on Web Intelligence, Mining and Semantics*, 1-8.

[5] Freeman, L. C. 1977. A set of measures of centrality based on betweenness. *Sociometry*, 35-41.

[6] Gutenbrunner, T., Leeds, D.D., Ross, S., Riad-Zaky, M., and Weiss, G.M. 2021. Measuring the academic impact of course sequencing using student grade data. In *Proc. of the 14$^{th}$ International Conference on Educational Data Mining*.

[7] Kleinberg, J. M. 1999. Authoritative sources in a hyperlinked environment. *Journal of the ACM (JACM)*, 46.5 (1999): 604-632.

[8] Leeds, D. D., Zhang, T., and Weiss, G. M. 2021. Mining course groupings using academic performance. In *Proceedings of the 14$^{th}$ International Conference on Educational Data Mining*.

[9] Marsden, P. V. 2005. *Encyclopedia of Social Measurement*.

[10] Page, L., Brin, S., Motwani, R., and Winograd, T. 1999. The PageRank citation ranking: Bringing order to the web. Stanford InfoLab.

[11] Park, H. W., and Thelwall, M. 2003. Hyperlink analyses of the World Wide Web: A review. *Journal of Computer-Mediated Communication*, 8(4).

[12] Riad-Zaky, M., Weiss, G.M., and Leeds, D.D. Course Grade Analytics with Networks (CGAN) [computer software], April 2021.

[13] Ruhnau, B. 2000. Eigenvector-centrality a node-centrality? *Social networks*, 22, 357–365.

[14] Weiss, G.M., Nguyen, N., Dominguez, K., and Leeds, D.D. 2021. Identifying hubs in undergraduate course networks based on scaled co-enrollments. In *Proceedings of the 14th International Conference on Educational Data Mining*.